\documentclass[twocolumn,aps,prl,10pt]{revtex4-1}
\usepackage{float}
\usepackage{amsmath}
\usepackage{amssymb}
\usepackage{graphicx}
\usepackage{braket}
\begin{document}
\title{Exact band structures for 1D superlattices   
       beyond the tight-binding approximation} 
\author{Olaf Kr\"uger and Alejandro Saenz}
\affiliation{Institut f\"ur Physik, Humboldt-Universit\"at zu Berlin, 
  Newtonstr.~15, 12489 Berlin, Germany}
\date{\today}
\begin{abstract}
  The band structures describing non-interacting particles in 
  one-dimensional superlattices of arbitrary periodicity are 
  obtained by an analytical diagonalization of the Hamiltonian  
  without adopting the popular tight-binding approximation.
  The results are compared with those of the tight-binding
  approximation. In this way, a quantitative prediction of the 
  validity and failure of the tight-binding approximation becomes 
  possible. In particular, it is demonstrated that in contrast to the 
  prediction of the tight-binding approximation the central energy
  bands do not touch for periodicities $\tau$ of the lattice where 
  $\tau=4n$ and $n$ is an integer.
\end{abstract}
\maketitle

%%%%%%%%%%%%%%%%
% Introduction %
%%%%%%%%%%%%%%%%
%
\textit{Introduction} --- The physics of optical lattices was studied 
intensely over the past ten years, starting with the first experimental 
realization in which ultracold atoms were trapped in the periodic potential 
generated by standing waves of laser light 
\cite{greiner200239,cold:lewe07,cold:bloc08}. They are 
still of experimental and theoretical interest due to the fact that 
both the lattice parameters (depth, spacing, and even overall 
geometry) and the atom-atom interaction can be experimentally tuned 
within large ranges. This permits the use of these systems as 
quantum simulators for, e.\,g., Hamiltonians of solid-state physics. 
Combined with the recently achieved single-site addressability 
(detection and manipulation) ultracold atoms (or molecules) in 
optical lattices are very promising candidates for a possible 
realization of quantum computers \cite{PhysRevLett.82.1975}. 
The usual approach adopted in the description of the physics in such 
uniform optical lattices is the Hubbard model using a tight-binding 
approximation (TBA) in which only next-neighbor hopping and a 
delta-function (on-site) interaction is considered. 

Optical superlattices \cite{cold:jaks98,cold:sebb06,cold:foel07} 
are a variation of the uniform lattices and are 
obtained by a superposition of at least two different standing waves
of laser light with different frequencies. The ratio $\tau$ of the 
adopted wavelengths equals the periodicity of the resulting 
superlattice. Such superlattices provide a possible realization of 
a quantum computer \cite{0953-4075-39-10-S19,cold:mazz12,cold:bloc12}. 
The Hamiltonian of non-interacting particles trapped in a superlattice was 
solved analytically within the TBA \cite{PhysRevB.73.174516}. 
Such solutions are of great interest, since they provide a good approximation 
for the band structure in the weakly interacting limit. In fact, it is 
now experimentally possible to reduce or even turn off the effective 
interaction between the particles using Feshbach resonances 
\cite{RevModPhys.82.1225}. Furthermore, especially analytical one-particle 
solutions can be the starting point for deriving perturbative or sometimes 
even exact results for interacting few- or even many-body systems. 
For example, exact solutions within the 
TBA for two interacting particles in a superlattice with periodicity 2 
were found with an extended Bethe {\it ansatz} \cite{cold:vali10}. 
Finally, part 
of the interest in one-dimensional systems stems from the fact that they 
provide the highest chances for finding analytical 
results \cite{gen:matt93,cold:caza11}. This is always 
very attractive from a general point of view and leads sometimes 
(via mapping) to analytical solutions of much more complicated 
problems.  

In this work the band structure for an optical superlattice with 
arbitrary periodicity is obtained without adopting the TBA. Thus 
it generalizes the results obtained recently within the 
TBA \cite{PhysRevB.73.174516}. While in the latter approximation 
the central energy bands were found to touch for all $\tau=4n$  
where $n$ is an integer, the rigorous results obtained beyond the 
TBA show that this behavior is in reality modified and the bands 
do not touch. Besides this important qualitative difference, the 
present results provide also the possibility to study quantitatively 
the failure of the TBA in optical superlattices for flat lattices.   

%%%%%%%%%%%%%%%
% Main Part 1 %
%%%%%%%%%%%%%%%
%
%
\begin{figure*}[t]
  \centering
  \includegraphics[width=0.30\textwidth]{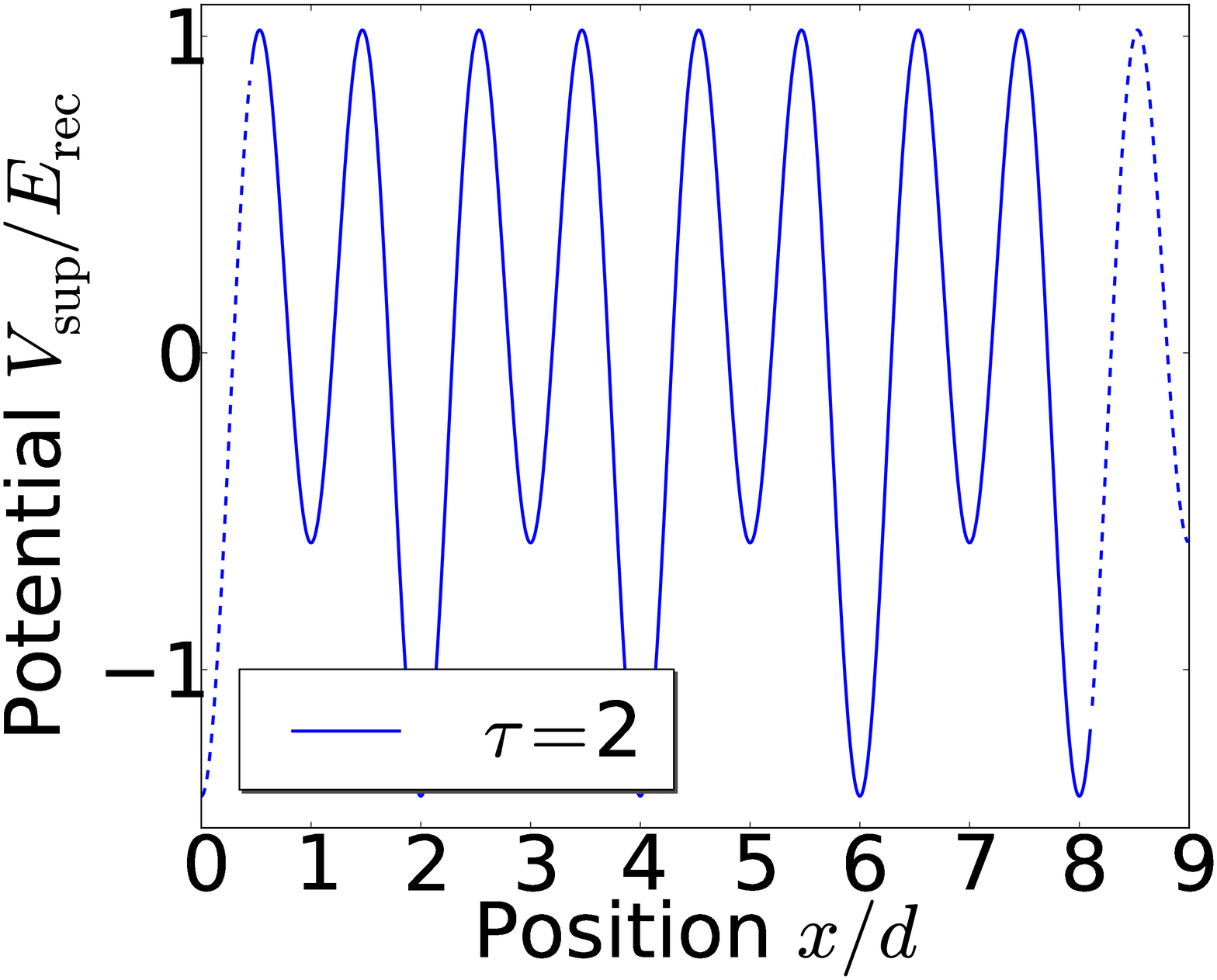}\label{fig:1_tau=2}%
  \hspace*{0.2cm}%
  \includegraphics[width=0.30\textwidth]{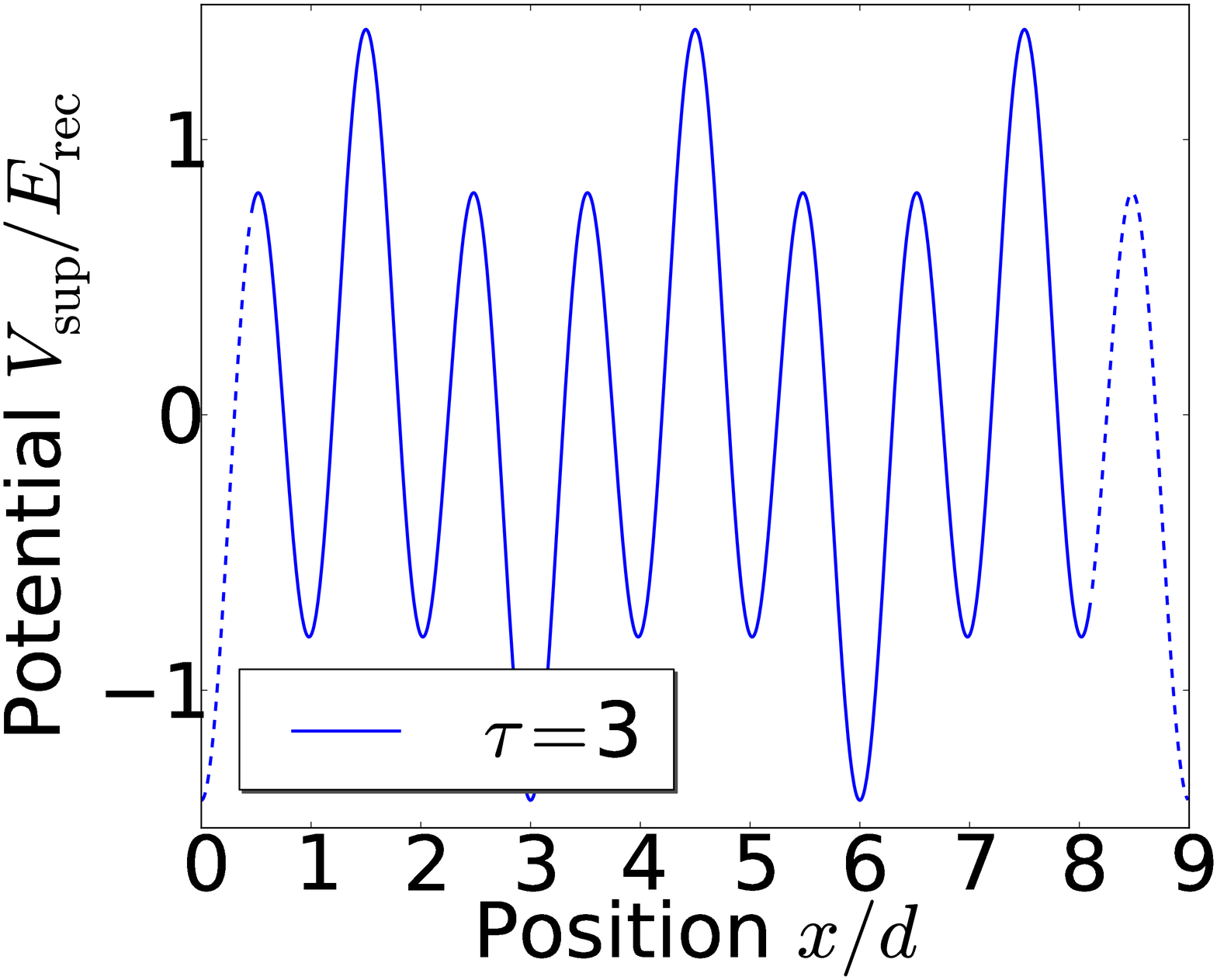}\label{fig:1_tau=3}%
  \hspace{0.2cm}%
  \includegraphics[width=0.30\textwidth]{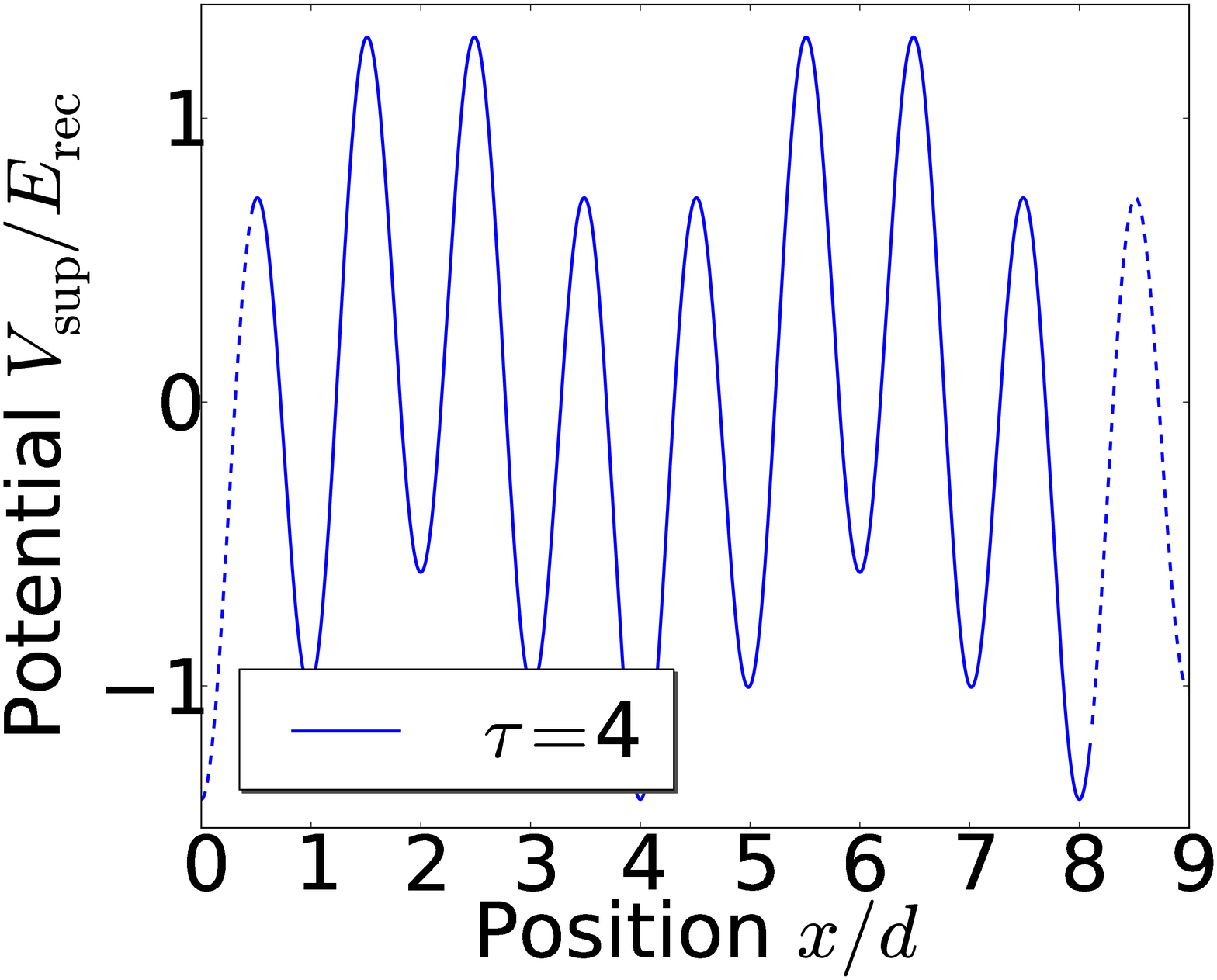}\label{fig:1_tau=4}
  \caption{(Color online) Example superlattice potentials 
    $V_{\mathrm{sup}}$ [Eq.~(\ref{eq:Vsup})] for 
    $V_0 = E_\mathrm{rec}, V_1 = 0.4E_\mathrm{rec}$, and different 
    periodicities $\tau$.}
  \label{fig:superlattices}
\end{figure*}
\textit{Hamiltonian} --- In this work, a one-dimensional
superlattice potential
\begin{equation}
\label{eq:Vsup}
    V_{\mathrm{sup}} = V_G(x) + V_M(x) = -V_0\cos(2\pi x) 
               - V_1\cos\left(\frac{2\pi x}{\tau}\right)
\end{equation}
is considered which is a superposition of a standard 
lattice potential $V_G(x)$ and a modulation lattice $V_M(x)$ with
periodicity $\tau$.  The problem to be solved is related to the 
Hill equation \cite{gen:hill86,gen:nist10}. 
To the authors' knowledge, only for the specific case of $\tau=2$ 
(Whittaker-Hill equation) a solution exists in literature, but even 
then only as an infinite 
Fourier-expansion with a recipe for calculating the coefficients.
Here and in the following all lengths are given 
in units of the lattice spacing $d$.
Hence, momenta are given in units of $d^{-1}$. The superlattice has
$N$ lattice sites at integer values $x$ ranging from $0$ to $N-1$.  
Periodic boundary conditions are used. Examples for various types of 
optical superlattices are depicted in Fig.~\ref{fig:superlattices}. 
The system is set up using low temperatures. Thus, there are $N$
one-particle states forming the Hilbert space $\mathcal{H}$ of the
system which is called the 1.~Bloch band. Without the modulation
lattice, $V_1=0$, the system's one-particle Hamiltonian would read
$\hat h_1 = -\frac{\hbar^2}{2M}\frac{\mathtt{d}^2}{\mathtt{d}x^2} -
V_0\cos(2\pi \hat x)$, where $M$ denotes the mass of the considered
particles. 
In the following, all energies are given in units of the
recoil energy $E_\mathrm{rec}=\frac{\pi^2\hbar^2}{2M}$. The
eigenstates $\ket{k}$ and eigenenergies $\epsilon(k)$ of $\hat h_1$
are analytically well known. $\ket{k}$ are called Bloch states.
Bloch's theorem states that $\Psi_k(x):=\Braket{x|k}=e^{ikx} u_k(x)$
where $u_k(x)$ is a periodic function in $x$. From this theorem and
periodic boundary conditions $\Psi_k(x+N) = \Psi_k(x)$ follows that
the $k$-quantum numbers, which represent the quasi momenta, are
discrete,
\begin{eqnarray}
  \label{eq:1}
  k = \frac{2\pi}{N}s, \quad s\in \mathbb{Z}.
\end{eqnarray}
Furthermore, the $N$ Bloch states $\ket{k}$ with $k \in [-\pi,+\pi)$
form a complete and orthonormal basis of $\mathcal{H}$.

\begin{figure}[ht]
  \centering
    \includegraphics[width=0.48\textwidth]{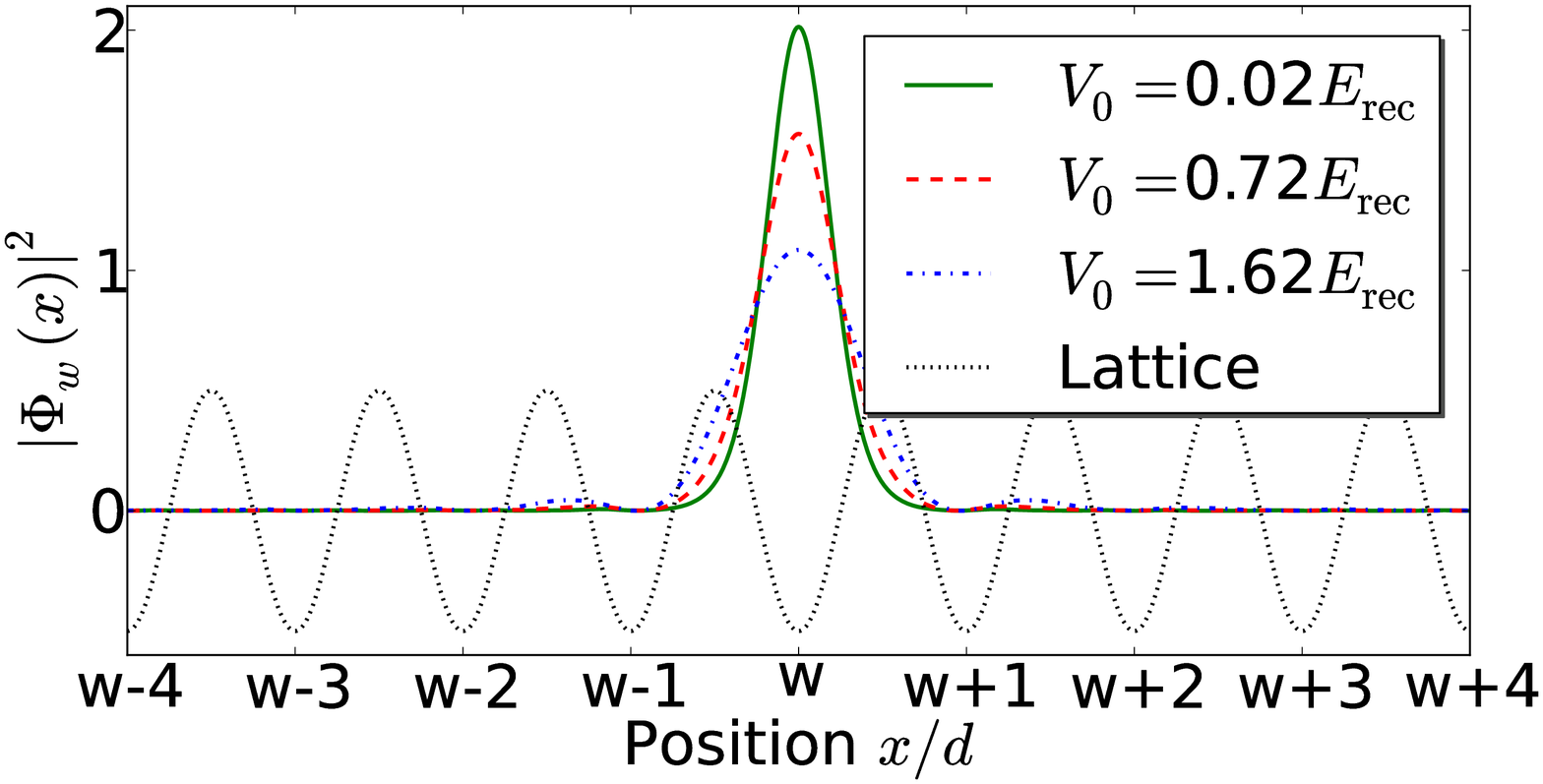}
    \caption{(Color online) Squared absolute values of the Wannier 
      wave functions $|\Phi_w(x)|^2$ for a uniform optical lattice 
      ($V_1=0$) with different $V_0$. For better readability, 
      only a single optical lattice with arbitrary amplitude is shown 
      (dotted line) to indicate the positions of the minima. 
%      For deeper latices the Wannier wave functions are more 
%      localized.
       }
  \label{fig:Wannierfunction}
\end{figure}

The Fourier transformation
\begin{eqnarray}
  \label{eq:2}
  \Ket{w}=\sum_{\ket{k}\in \mathcal{H}} \sqrt{\frac{1}{N}} e^{-ikw}\Ket{k}
\end{eqnarray}
defines the Wannier states $\ket{w}$ which also form a complete and
orthonormal basis of $\mathcal{H}$ choosing $w$ in between $0$ and
$N-1$. The absolute value square of the Wannier wave function
$\Phi_w(x)=\Braket{x|w}$ is depicted in Fig.~\ref{fig:Wannierfunction}.
As can be seen, $\Phi_w(x)$ is well located at the lattice site $w$, 
especially for deep latices. From Eqs.~(\ref{eq:1}) and (\ref{eq:2})
follow periodic boundary conditions for the Wannier states,
\begin{eqnarray}
  \label{eq:3}
  \Ket{w+N}=\Ket{w}.
\end{eqnarray}
In this basis, the Hamiltonian of the system is
\begin{eqnarray}
  \label{eq:4}
  \hat H=\sum_{ww'}\mathcal{J}\left(w,w'\right) \hat a^\dagger_w \hat a_{w'},
\end{eqnarray}
with the matrix elements
\begin{eqnarray}
  \label{eq:5}
  \mathcal{J}\left(w,w'\right) &=&\Braket{w|\hat T + \hat V_G(\hat
    x) + \hat V_M(\hat
    x)|w'}\nonumber\\
  &=& \sum_k \frac{\epsilon(k)}{N} e^{ik(w-w')} - \nonumber\\
  &&- V_1 \int \textrm{d}x \cos\left(\frac{2\pi}{\tau} x\right) \Phi^*_w(x) \Phi_{w'}(x)
\end{eqnarray}
and the annihilation and creation operators of Wannier states 
$\hat a_w$ and $\hat a^\dagger_w$, respectively, that fulfill the usual bosonic 
(fermionic) (anti-)commutator relations.

The TBA neglects all $\mathcal{J} \left( w,w' \right)$ with $|w-w'|>1$
and assumes a small overlap between different Wannier functions, $ \left[
  \Phi_w^*(x)\Phi_{w'}(x) \right]_\mathrm{TBA} =
\delta_{ww'}\delta(w-x)$. Then, the chemical potential $\mu_0:=\sum_k
\frac{\epsilon(k)}{N}$ and the Hopping parameter $J := - \sum_k
\frac{\epsilon(k)}{N}e^{ik}$ can be defined and the Hamiltonian in TBA
reads
\begin{eqnarray}
  \label{eq:6}
  \hat H=& \: \sum_{w}\left[ -J \cdot \hat a^\dagger_w \hat a_{w-1} +
    \text{h.c.} \right] + \mu_0 \sum_{w}\hat a^\dagger_w \hat a_w - \nonumber\\ &- V_1
  \sum_w \cos\left(\frac{2\pi}{\tau} w\right) \hat a^\dagger_w \hat a_{w}.
\end{eqnarray}

%%%%%%%%%%%%%%%
% Main Part 2 %
%%%%%%%%%%%%%%%
% 
\textit{Solutions of the Schr{\"o}dinger equation} --- As depicted in 
Fig.~\ref{fig:superlattices}, the lattice is invariant under a translation
of $\tau$ lattice sites. Thus, the 1.~Bloch band can be divided in
$\tau$ subspaces $\mathcal{H} = \mathcal{H}_0 \oplus \mathcal{H}_1
\oplus \ldots \oplus \mathcal{H}_{\tau-1}$ so that $\Ket{w = l\tau +
  \overline{m}} =: \Ket{l,m} \in \mathcal{H}_{\overline{m}}$, where
$\overline z = z \mod \tau$ for any integer $z$. Then, creators and
annihilators can be written as
\begin{eqnarray}
  \label{eq:7}
  \hat a^\dagger_m(l)=\hat a^\dagger_{w=l \tau
    +\overline{m}}, \qquad \hat a_m(l)=\hat a_{w=l \tau +\overline{m}} 
\end{eqnarray}
and the system's Hamiltonian (\ref{eq:4}) reads
\begin{eqnarray}
  \label{eq:8}
  \hat H&=&\sum_{m,\Delta m=0}^{\tau-1} \sum_{l,\Delta
    l=0}^{\frac{N}{\tau}-1} \mathcal{J}\left((l+\Delta
    l)\tau+\overline{m+\Delta m},l\tau+\overline{m} \right)\nonumber\\
  && \qquad \times \; \hat a^\dagger_{m+\Delta m}(l+\Delta l) \hat a_m(l).
\end{eqnarray}
As it turns out, the $\mathcal{J} \left( (l+\Delta l) \tau + \overline{
    m+\Delta m } , l \tau + \overline{m} \right)$ do not depend on the
quantum number $l$.

\begin{figure}[t]
  \centering
  \includegraphics[width=0.48\textwidth]{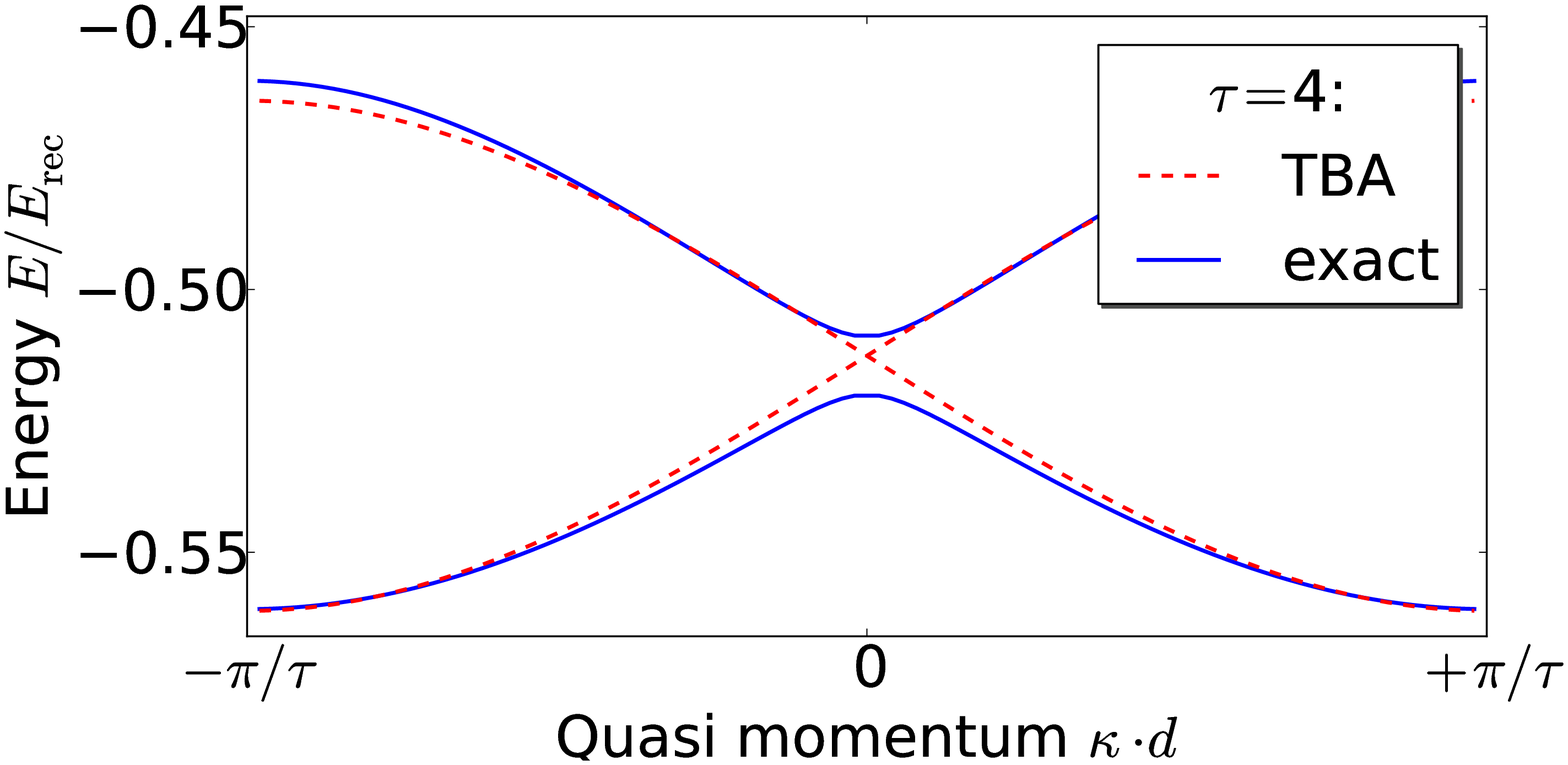}
  \includegraphics[width=0.48\textwidth]{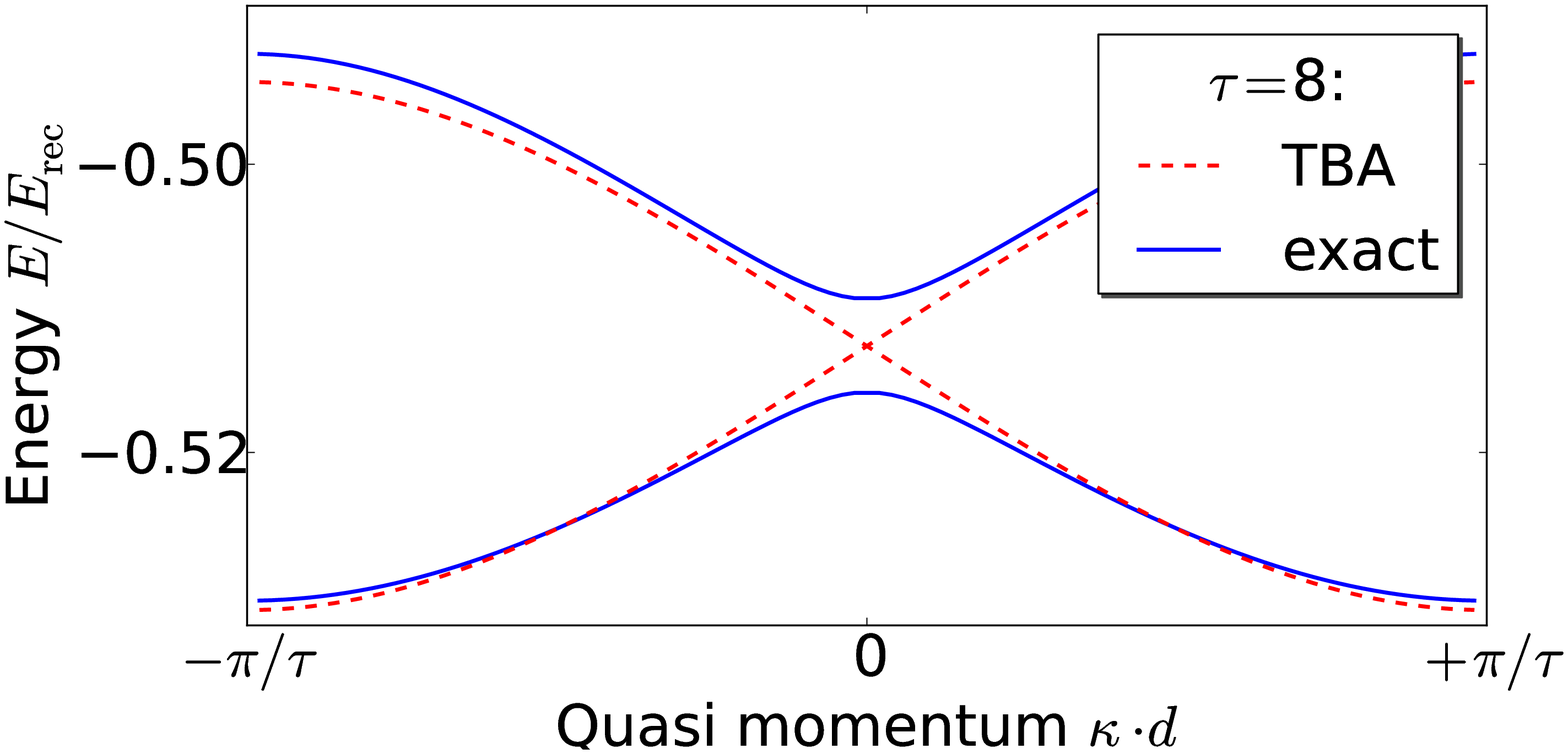} 
  \caption{(Color online) Central energy bands for an optical
    superlattice $V_{\mathrm{sup}}$ [see Eq.~(\ref{eq:Vsup})]
    with $V_0 = 2.42E_\mathrm{rec}$ and $V_1 = 0.06 V_0$ for 
    $\tau=4$ and $\tau=8$. In the 
    tight-binding approximation (TBA---red dashed line) these energy 
    bands touch whereas a gap is found in the 
    exact calculation (exact---blue solid line).}
  \label{fig:result1}
\end{figure}

Fourier-transformed states can be defined as
\begin{eqnarray}
  \label{eq:9}
  \Ket{\kappa ,m}:=\sum_{l=0}^{\frac{N}{\tau}-1} \sqrt{\frac{\tau}{N}}
  e^{i\kappa \left(l\tau+\overline{m}\right)}
  \Ket{l,m},
\end{eqnarray}
where those $\Ket{\kappa,m}$ with $\kappa = \frac{2\pi}{N}s$, $s\in
\mathbb{Z}$, $\kappa \in\left[-\frac{\pi}{\tau} , \frac{\pi}{\tau}
\right)$ form an orthonormal basis of the Hilbert subspace known as 
\textit{first Brillouin zone} (\textbf{1.\,BZ}). Then, the bosonic 
(fermionic) annihilators and creators of those Fourier-transformed 
states are given by 
\begin{equation}
  \label{eq:10}
  \hat b_m(\kappa ) = \sum_{l=0}^{\frac{N}{\tau}-1}
  \sqrt{\frac{\tau}{N}} e^{-i\kappa \left(l\tau+\overline{m}\right)}
  \hat a_m(l)
\end{equation} 
and its adjoint equation. They 
fulfill bosonic (fermionic) (anti-)commutator relations
respectively. Introducing them into the Hamiltonian (Eq.~(\ref{eq:8})) and
rewriting $\alpha = \overline{m+\Delta m}$, $\beta = \overline{m}$ gives
\begin{eqnarray}
  \label{eq:11}
  \hat H &=& \sum_{\kappa\in\textbf{1.BZ}}
  \sum_{\alpha,\beta=0}^{\tau-1} \sum_{\Delta l=0}^{\frac{N}{\tau}-1} 
  \mathcal{J}\left((l+\Delta l)\tau+\alpha,l\tau+\beta
  \right)\times\nonumber\\
  &&\times e^{-i\kappa\left(\Delta l\tau+\alpha - \beta\right)} \hat
  b^\dagger_{\alpha}(\kappa ) \hat b_\beta(\kappa)\nonumber\\
  &=& \sum_{\kappa\in\textbf{1.BZ}}
  \sum_{\alpha,\beta=0}^{\tau-1} \hat
  b^\dagger_{\alpha}(\kappa ) \mathcal{M}_{\alpha\beta} \hat b_\beta(\kappa),
\end{eqnarray}
where the elements of the introduced matrix $\mathcal{M}$ can be
calculated using Eq.~(\ref{eq:5}) and performing the sum over $\Delta
l$,
\begin{eqnarray}
  \label{eq:12}
  \mathcal{M}_{\alpha\beta} &=& \sum_{\Delta l=0}^{\frac{N}{\tau}-1} 
  \mathcal{J}\left((l+\Delta l)\tau+\alpha,l\tau+\beta
  \right) e^{-i\kappa\left(\Delta l\tau+\alpha -
      \beta\right)}\nonumber\\
  &=& \frac{1}{\tau}\sum_k^{\kappa\star} e^{i(k-\kappa)\left(\alpha -
      \beta \right)} \Bigg(\epsilon(k) - V_1
  \frac{N}{\tau}\sum_{k'}^{k\star}e^{i\beta
    \left(k-k'\right)}\times\nonumber\\ 
  &&\times\int_{0}^{\tau} \textrm{d}y \cos\left(\frac{2\pi}{\tau} y\right)
  \Psi_k^*(y) \Psi_{k'}(y)\Bigg).
\end{eqnarray}
Here, $\sum_k^{q\star}$ denotes a sum over those quantum numbers
$k\in[-\pi,\pi)$, which fulfill Eq.~(\ref{eq:1}) and the condition 
$k-q = \frac{2\pi}{\tau}$, $s\in\mathbb{Z}$ for any value $q$. As 
can be shown, $\mathcal{M}$ is hermitian and thus, has real eigenvalues
$E_0\ldots E_{\tau-1}$. The columns of the hermitian matrix
$\mathcal{U}$, which diagonalizes $\mathcal{M}$,
$\mathcal{U}\mathcal{M}\mathcal{U}^\dagger = \mathrm{diag} \left(
  E_0,E_1,\ldots,E_{\tau-1} \right)$, are given by the eigenvectors of
$\mathcal{M}$. Finally, the creators and annihilators defined by
\begin{equation}
  \label{eq:13}
  \left( \hat\psi^\dagger_0(\kappa), \ldots,
    \hat\psi^\dagger_{\tau-1}(\kappa) \right) = \left(\hat 
    b^\dagger_0(\kappa),\ldots,\hat
    b^\dagger_{\tau-1}(\kappa)\right)\mathcal{U}^\dagger
\end{equation}
and its adjoint eqution fulfill the bosonic (fermionic) (anti-)commutator 
relations respectively. Introducing them into Eq.~(\ref{eq:11}) finally gives
the diagonal Hamiltonian
\begin{eqnarray}
  \label{eq:14}
  \hat H = \sum_{\kappa\in\textbf{1.BZ}} \, \sum_{\alpha=0}^{\tau-1}
  E_\alpha(\kappa) \,\hat\psi_\alpha^\dagger(\kappa)\,
  \hat\psi_\alpha(\kappa).
\end{eqnarray}
Then, its $N$-particle bosonic or fermionic eigenstates are given in
Fock space by
\begin{equation}
  \label{eq:15}
  \Ket{n_{\alpha_1}(\kappa_1), \ldots, n_{\alpha_N}(\kappa_N) } =
  \prod_{i=1}^N\frac{\left(\hat\psi_{\alpha_i}^\dagger(\kappa_i)
    \right)^{n_{\alpha_i}( \kappa_i)}}{\sqrt{n_{ \alpha_i}(
      \kappa_i)!}}\Ket{\emptyset}
\end{equation}
for $\alpha_i\in[0,\tau-1]$, $\kappa_i\in\mathbf{1.BZ}$, and where
$\Ket{\emptyset}$ denotes the vacuum state. The eigenenergies
$E_\alpha(\kappa)$ are given by the eigenvalues of $\mathcal{M}$. 
They can be obtained easily, since the matrix elements 
$\mathcal{M}_{\alpha\beta}$ in 
Eq.~(\ref{eq:12}) can be calculated analytically (see Supplemental 
Material).

%%%%%%%%%%%
% Results %
%%%%%%%%%%%

\begin{figure}[t]
  \centering
  \includegraphics[width=0.48\textwidth]{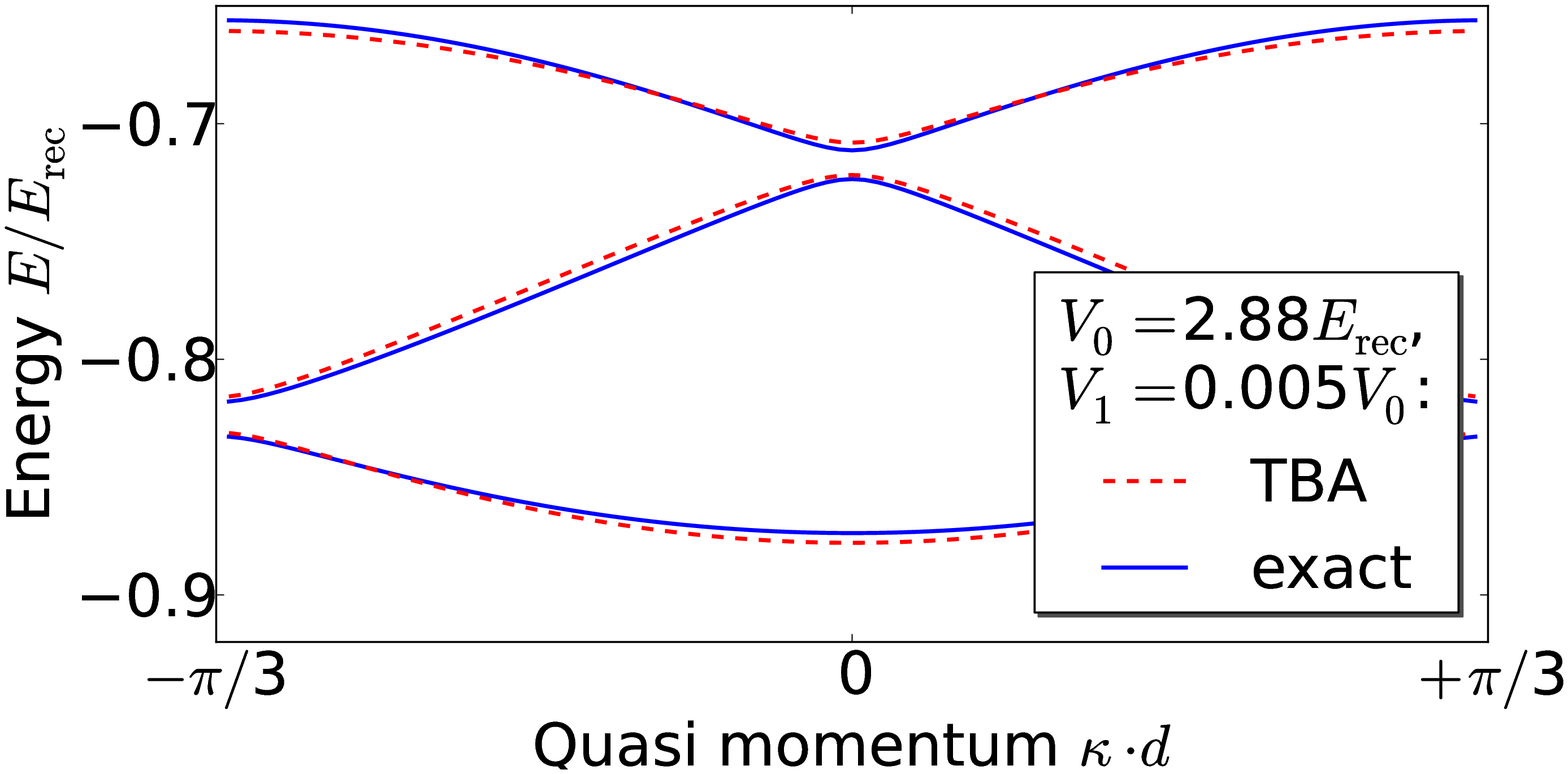}
  \includegraphics[width=0.48\textwidth]{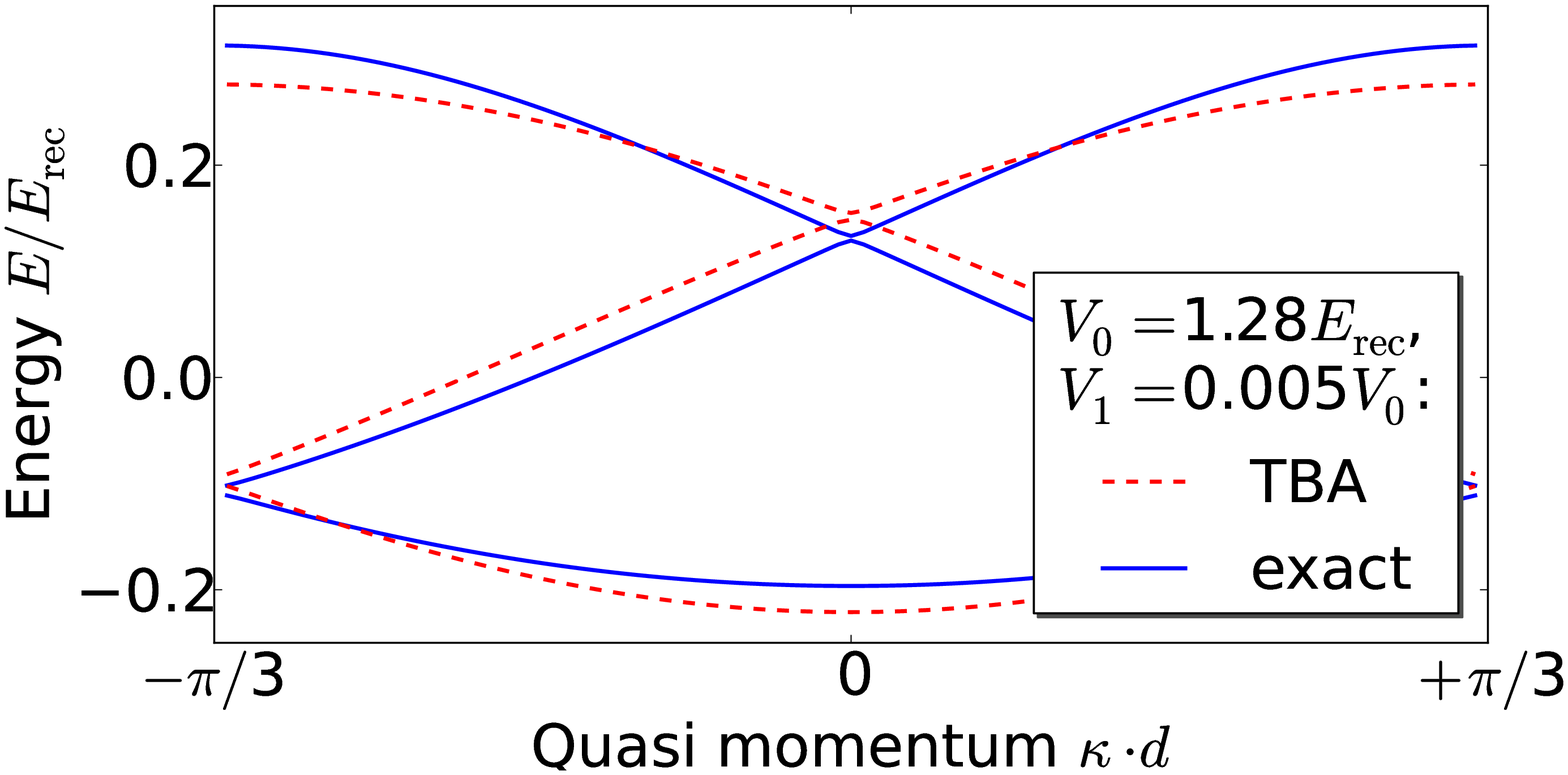}
  \includegraphics[width=0.48\textwidth]{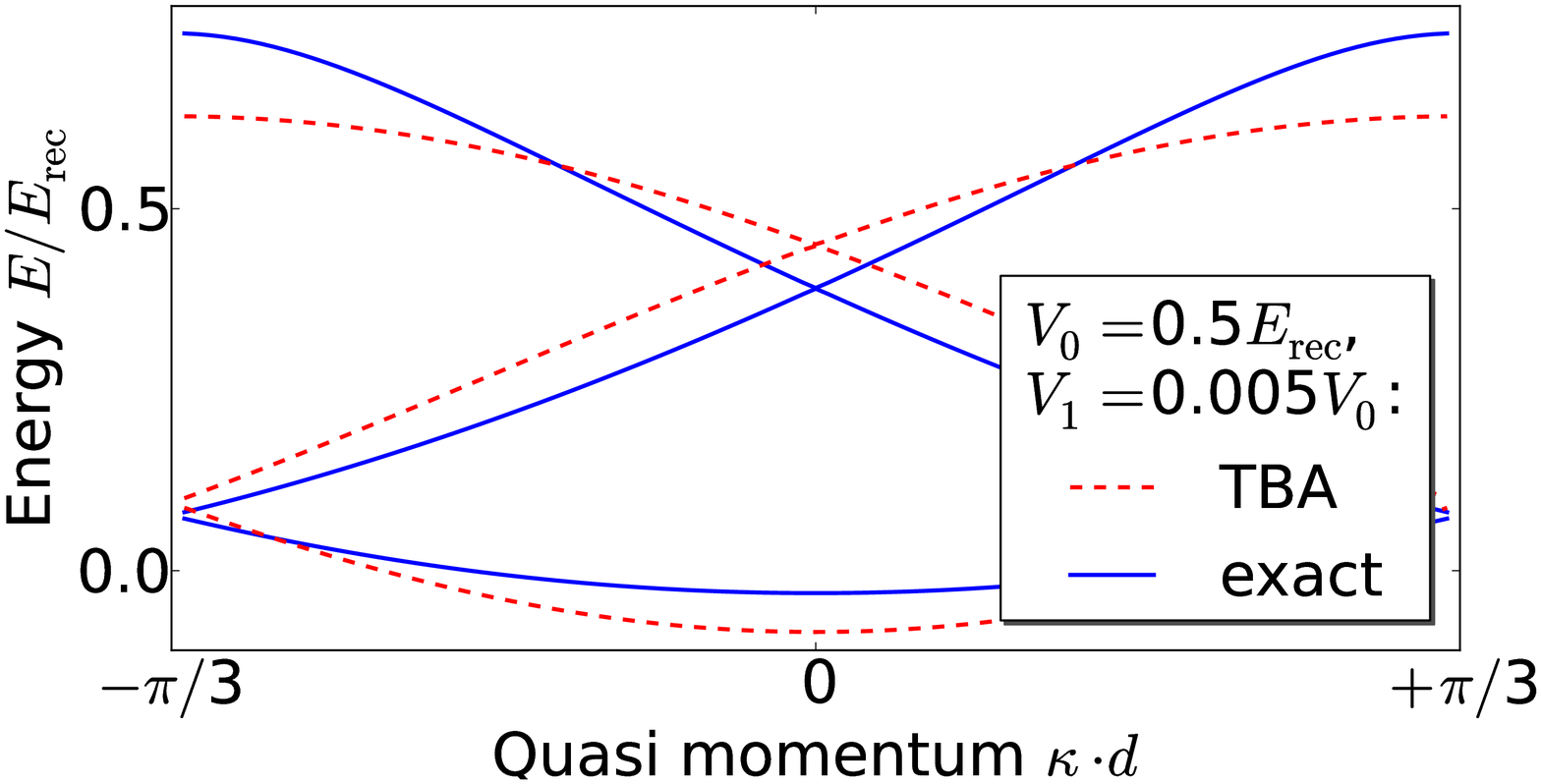}
  \caption{(Color online) Energy bands (TBA: red dashed, 
    exact: blue solid) for an optical superlattice 
    $V_{\mathrm{sup}}$ [see Eq.~(\ref{eq:Vsup})]
    with periodicity $\tau=3$ for different values of the 
    lattice depth $V_0$ (specified in the graphs) 
    and the modulation $V_1=0.005V_0$.}
  \label{fig:result2}
\end{figure}
\begin{figure}[t]
  \centering
  \includegraphics[width=0.48\textwidth]{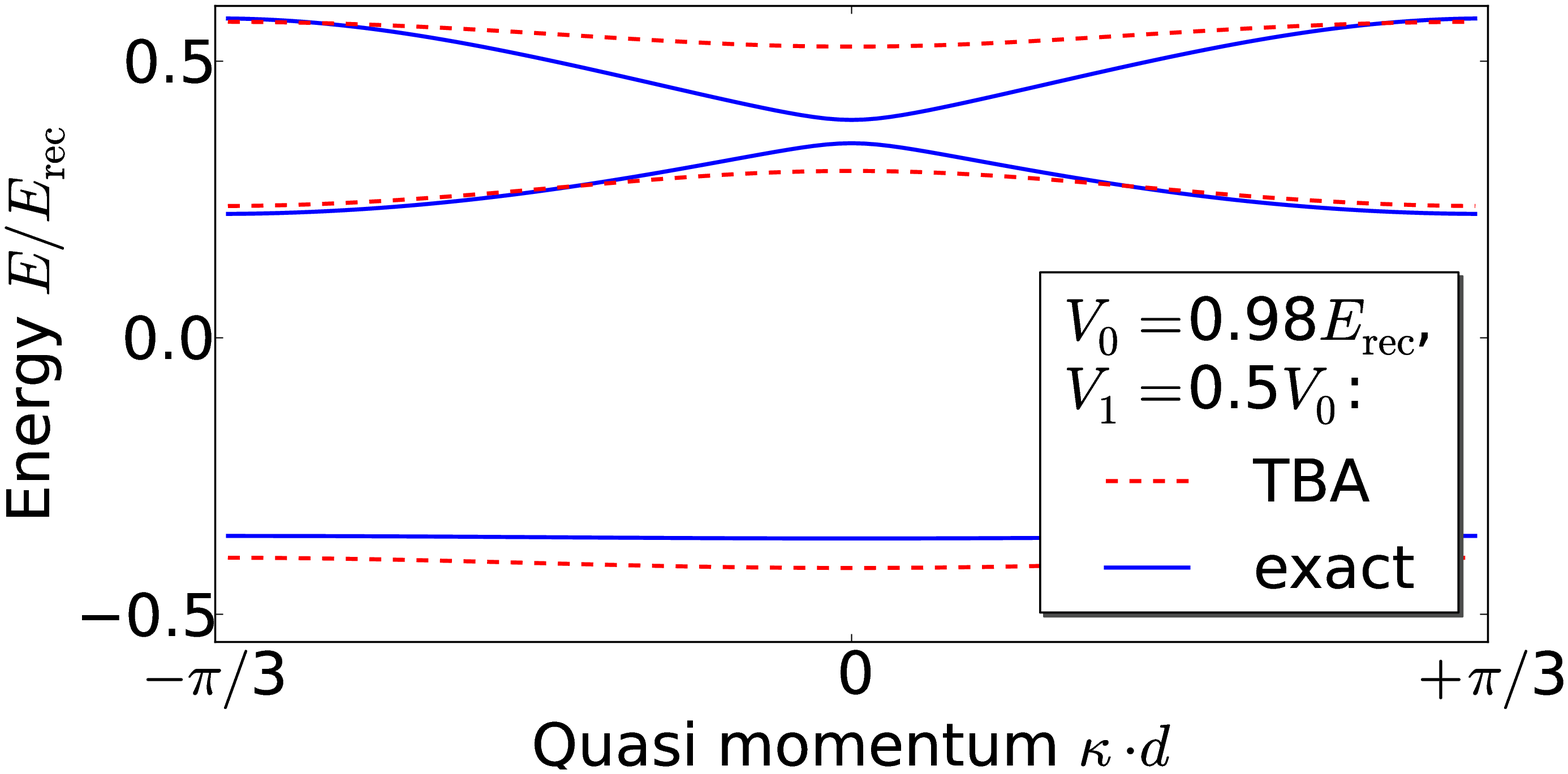}
  \includegraphics[width=0.48\textwidth]{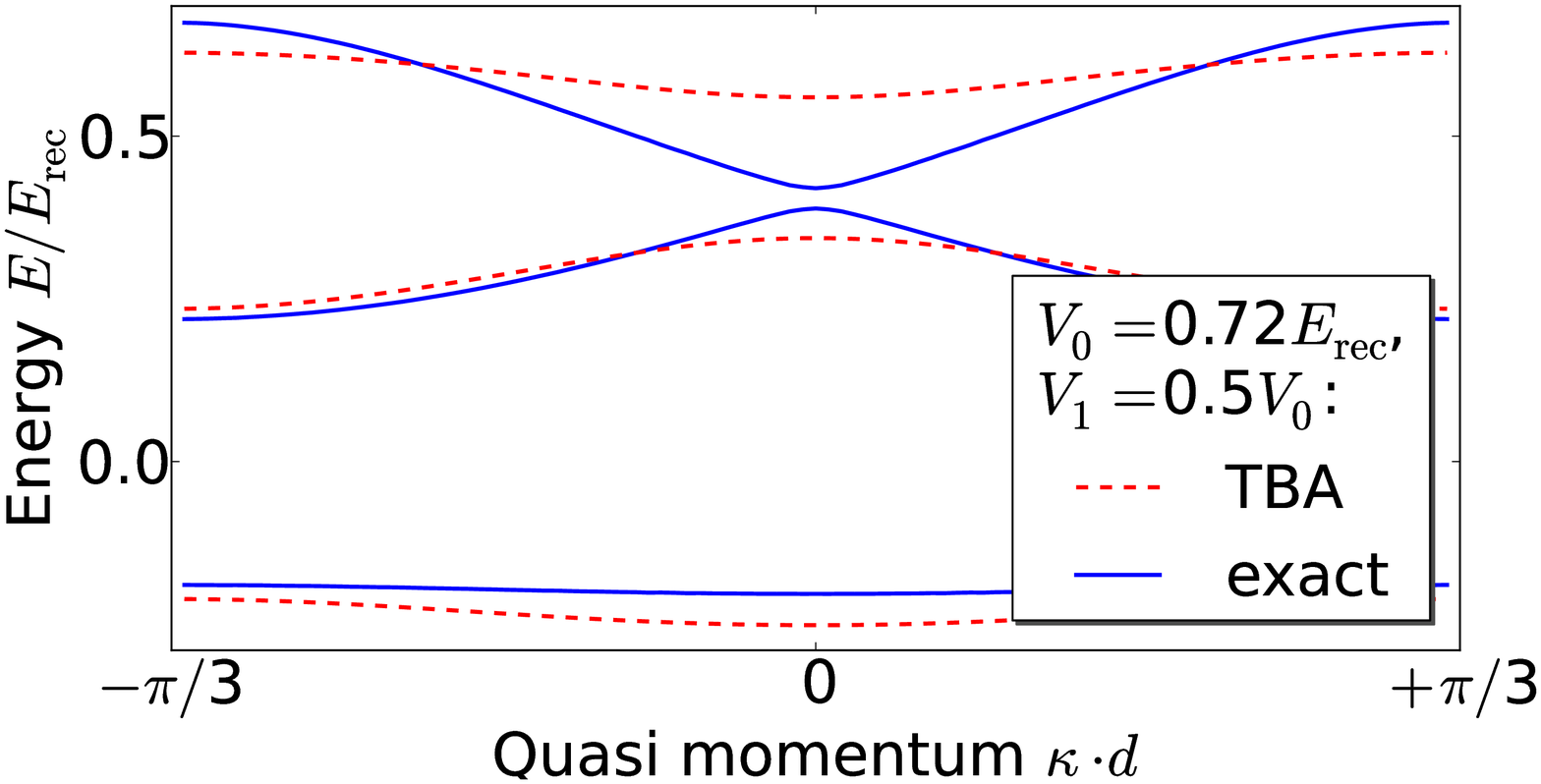}
  \caption{(Color online) as Fig.~\ref{fig:result2}, 
    but for $V_0=0.98\,E_{\mathrm{rec}}$ (top) and 
    $V_0=0.72\,E_{\mathrm{rec}}$ (bottom) and a much stronger modulation 
    potential $V_1 = 0.5V_0$.}
  \label{fig:result3}
\end{figure}

\textit{Comparison to the TBA results} --- After the main result of 
this work, the exact diagonalization of the Hamiltonian, was presented, 
some implications are discussed. Whenever $\tau=4n$, $n$ being an
integer, the TBA predicts the central energy bands to touch at
$\kappa=0$ \cite{PhysRevB.73.174516}. However, as can be seen from 
the exemplary cases $\tau=4$ and $\tau=8$ in Fig.~\ref{fig:result1}, 
this contact disappears in the exact calculation. Therefore, the 
contact is an artifact of the TBA. As an important consequence, 
the in \cite{PhysRevB.73.174516} predicted non-insulating properties 
for superlattices with $\tau=4n$ and half filling of the states 
does not appear in reality. Besides this qualitative difference, 
there are, of course, also quantitative ones. As an example, 
Fig.~\ref{fig:result2} shows the energy bands for a $\tau=3$ 
superlattice with different depths $V_0$. Clearly, for deep 
lattices ($V_0 \gtrsim 2.88$), a very good agreement between the 
TBA and the exact results is found, whereas for decreasing lattice 
depths ($V_0 \lesssim 1.28$) the agreement becomes worse, as one 
would expect. Of course, only the exact results allow for a 
quantitative measure of the breakdown of the TBA. While for 
very weak modulation potentials $V_1$ ($V_1=0.005\,V_0$ in 
Fig.~\ref{fig:result2}) the deviations of the TBA bands from 
the correct ones are relatively uniformly distributed over the 
quasi-momenta, this is not the case for larger modulation potentials. 
As can be seen in Fig.~\ref{fig:result3} ($V_1=0.5\,V_0$), the dispersion 
of the bands differs in a pronounced fashion between the TBA and the exact 
results. As a consequence, the gap is largely overestimated by the TBA. 
In fact, the TBA predicts almost discrete energy
levels whereas the uppermost energy bands of the exact calculation
are substantially broadened in the exemplary case $\tau=3$.

\textit{Conclusion} --- As is shown in this work, the Hamiltonian of 
non-interacting particles in a general one-dimensional superlattice 
can be solved exactly without invoking the TBA as is done in the 
popular Hubbard models. The only assumption introduced is the 
restriction of the Hilbert space to the 1.~Bloch band. 
In this case the problem can be reduced to the 
diagonalization of a square matrix with its dimension given by the 
periodicity $\tau$ of the superlattice and all matrix elements 
can be calculated exactly. Especially 
for periodicities $\tau=2$ or 3 fully analytical expressions can 
be found in principle and the solution includes the case of a uniform 
lattice ($V_1=0$). Thus, it is proven that the Schr{\"o}dinger 
equation of non-interacting particles in a 1D superlattice, and thus 
of a problem that is of great interest in view of current 
experiments with ultracold atoms, belongs to the small class of 
exactly integrable quantum-mechanical problems. 
It is also possible to extend the indicated method to higher
dimensions via a separation {\it ansatz} or to different 
translation-invariant lattice geometries such as graphene which has 
been studied within a tight-binding approximation 
in \cite{PhysRevLett.53.2449}. Analogously to the other 
cases for which analytical solutions of the independent-particle 
problems (fully or within, e.\,g., the Hubbard model) exist, 
the present results can be the starting point for perturbative, 
numerical, or possibly even analytical solutions for the 
corresponding problem of interacting particles. Of course, 
as is demonstrated explicitly in this work, the exact solutions 
are useful for qualitatively and quantitatively determining the 
validity or failure of the popular tight-binding approximation.  
%
%
%%%%%%%%%%%%%%%%
% Bibliography %
%%%%%%%%%%%%%%%%

\bibstyle{apsrev}
%\bibliography{literatur}
%
%
% 
\end{document}